\documentclass[10pt,prd,aps,twocolumn,preprintnumbers, showpacs, nofootinbib,superscriptaddress,notitlepage]{revtex4-2}
\usepackage{amssymb,amsthm,amsmath}
\usepackage{textcomp}
\usepackage{subfig,graphicx}   % figures
\usepackage{color}      % color is used in text
\usepackage{slashed}    % Feynman slash
\usepackage{verbatim}
\usepackage[normalem]{ulem}
\usepackage{rotating}   % to rotate tables
\usepackage{multirow}  
\usepackage[colorlinks,
linkcolor=red,
anchorcolor=blue,
citecolor=blue
]{hyperref}

\begin{document}

	\baselineskip=15pt
	
	\preprint{CTPU-PTC-26-01}

\title{
%{\color{blue}  Decoding CPV Hierarchy Amplitudes for Charmed Baryon Decays}\\
  Decoding Weak Phase Amplitudes in Charmed Baryon Decays
}

\affiliation{ Department of Physics and Institute of Theoretical Physics, Nanjing Normal University, Nanjing, Jiangsu 210023, China}
\affiliation{Particle Theory and Cosmology Group, Center for Theoretical Physics of the Universe, Institute for Basic Science (IBS), Daejeon 34126, Korea }
\affiliation{State Key Laboratory of Dark Matter Physics, Tsung-Dao Lee Institute and School of
Physics and Astronomy, Shanghai Jiao Tong University, Shanghai 201210, China}
\author{Zhi-Peng  Xing$^{1}$}
%\email{zpxing@nnu.edu.cn} 
\author{Jin Sun$^{2}$}
\email{sunjin0810@ibs.re.kr(Contact author)}
\author{Xiao-Gang He$^{3}$}
%\email{hexg@sjtu.edu.cn}

\begin{abstract}

In this Letter, we present a strategy to decode decay amplitudes with different weak phases in two-body charmed baryon decays, addressing a long-standing challenge to understanding CP violation in this sector.
Using SU(3) flavor symmetry, we disentangle the amplitudes defined through $M=\lambda_s A_s + \lambda_b A_b$ ($\lambda_i = V_{ui}V_{ci}^*$), enabling a data-driven determination of their relative hierarchy across Cabibbo-favored, singly, and doubly Cabibbo-suppressed modes. 
Applying this framework to current data, we find that the ratio 
$A_b/A_s$ is approximately 100 times larger than expectation of $\mathcal{O}(1)$ at $2.1\sigma$ level. 
This unexpected pattern may be accommodated by SU(3) breaking effects, but it can also point to new physics.
We further show that the Lee-Yang parameters provide an independent and complementary probe.
To resolve this ambiguity in the origin of the enhancement, we propose a decisive discriminator using decay modes insensitive to large SU(3) breaking, in particular $\Xi_c^0 \to pK^-$ and $\Xi_c^0 \to \Sigma^+\pi^-$ as golden channels. 
Measurements of these modes will distinguish between symmetry breaking effects and new physics, establishing a clean pathway to probe CP violation in the charm baryon sector.

\end{abstract}

\maketitle

\noindent{\bf \textit{Introduction}}
%big picture and motivation
Charge–Parity violation (CPV) is a crucial ingredient for understanding the matter–antimatter asymmetry of the universe. In the Standard Model (SM), CPV in charm decays is expected to be tiny due to the  Cabibbo-Kobayashi-Maskawa(CKM) suppression. However, this picture has been challenged by  the LHCb Collaboration~\cite{LHCb:2019hro,LHCb:2022lry}, which reported the larger-than-expected  CPV in neutral $D$-meson decays. This raises a natural question of whether  similar effects may also arise in the charmed baryon sector.

 % Why charmed baryons matter
 Two-body charmed-baryon decays provide a particularly clean and largely unexplored arena for addressing this question.
 For the modes $B_c \to B P$, where $B_c$ denotes the charm baryon, B the light baryon and P the meson, 
 the decay amplitudes exhibit a rich structure~\cite{Korner:1978tc,Cheng:1991sn,Lu:2016ogy,Geng:2017mxn,Li:2025nzx}. These channels are organized into Cabibbo-favored (CF), singly Cabibbo-suppressed (SCS), and doubly Cabibbo-suppressed (DCS) classes, offering a natural framework for separating contributions with different CKM hierarchies. 
 Each decay receives  parity-conserving and parity-violating amplitudes, allowing  the  decay matrix element $\mathcal{M}$ and  width $\Gamma$ as~\cite{PDG}
\begin{eqnarray} \label{ampli}
    &&\mathcal{M} = i \bar u_B (F - G \gamma_5) u_{B_c},\quad 
  |M|^2=|F|^2 + \kappa^2 |G|^2,   \nonumber \\
    &&\Gamma=\frac{p_B}{8\pi}\frac{(m_{B_c}+m_B)^2-m_P^2}{m_{B_c}^2} |M|^2. 
   % \left(|F|^2+\kappa^2|G|^2\right)   
\end{eqnarray}
Here $\kappa=p_B/(E_B+m_B)$, $E_B$ and $p_B$ are the energy and momentum  of $B$ in the rest frame of $B_c$.
Throughout this work, $F$ and $G$ denote the full amplitude structures, 
including both the purely hadronic contributions and the corresponding CKM factors.
For a final-state pseudoscalar meson, its negative intrinsic parity implies that 
 scalar term $F$ describes the parity-violating S-wave contribution, whereas the $\gamma_5$
 term $G$ describes the parity-conserving P-wave contribution.

% Physics mechanism and  hierarchy
Observable CPV arises only in SCS transitions, where amplitudes with different weak phases can interfere.
The dominant tree-level amplitudes scale with $\lambda_d = V_{cd}^\ast V_{ud}$ or $\lambda_s = V_{cs}^\ast V_{us}$, while penguin amplitudes are proportional to the much smaller combination $\lambda_b = V_{cb}^\ast V_{ub}$. 
Using the unitarity relation  $\lambda_d=-\lambda_s-\lambda_b$, 
the two parity amplitudes for SCS can be parameterized as 
$F=\lambda_s A_s^f+\lambda_b A_b^f$ and $G=\lambda_s A_s^g+\lambda_b A_b^g$, where
the superscripts $f,g$ label the F- and G-type parity structures, respectively, 
while the subscripts $s,b$ indicate the  hadronic components associated  with corresponding CKM factors.
In particular, $A_b$ receives not only penguin contributions but also specific tree components after the CKM rearrangement.
Since  both $A_s$ and  $A_b$  receive the tree level  contributions with  comparable Wilson coefficients, one naively expects $A_b/A_s\sim \mathcal{O}(1)$, 
implying  $|\lambda_b A_b/\lambda_sA_s|\sim O(10^{-3})$.
 Interference among them can generate observable CPV, making charmed baryons a sensitive probe of both hadronic dynamics and possible new physics~\cite{Xing:2024nvg,Yang:2025orn}.

In this Letter, we develop a data-driven SU(3) strategy that combines CF, SCS, and DCS modes, to separate $A_b$ and $A_s$  without relying on direct CPV measurements.
Current data favor $A_b/A_s=\mathcal{O}(100)$,
a $2.1\sigma$  departure from the naive  O(1) expectation.
We examine whether this enhancement can arise from $\mathcal{O}(10\%)$  
SU(3) breaking effects, 
and identify $\Xi_c^0 \to pK^-$ and $\Xi_c^0 \to \Sigma^+\pi^-$  as breaking insensitive 
channels that discriminate symmetry breaking from new physics.
\\

\noindent{\bf \textit{Charmed baryon decay amplitudes}}
We employ the irreducible representation amplitudes (IRA) of  SU(3) flavor symmetry for
anti-triplet charmed baryons $B_c$ decays into   an octet baryon $B$ and a pseudoscalar meson $P$.
The two light quarks in a ground-state charmed baryon form an anti-triplet $\overline{\mathbf{3}}$ and an sextet $\mathbf{6}$ under $\mathbf{3}\otimes\mathbf{3}=\mathbf{6}\oplus\overline{\mathbf{3}}$
The initial anti-triplet is indicated by $(\mathbf B_c)^{ij}$  or equivalently $(\mathbf B_c)_k=\frac{1}{2}\epsilon_{kij}(\mathbf B_c)^{ij}=(\Xi_c^0,-\Xi_c^+,\Lambda_c^+)$.
 The final states comprise 
 baryon octet $B:(n,p,\Lambda^0,\Sigma^{\pm,0},\Xi^{-,0})$,  meson octet $P_8:(\pi^{\pm,0},K^{\pm,0},\bar{K}^0,\eta_{8})$ and singlet $\eta_1$, as detailed in~\cite{Xing:2024nvg,He:2024unv}.
 The effective Lagrangian contains $3$, $\bar 6$ and $15$ components, with CF, SCS, and DCS parts distinguished by their CKM factors~\cite{Yang:2025orn}
\begin{eqnarray}
&&\mathcal{L}_{\mathrm{eff}}^{\mathrm{CF}}=-\frac{G_F}{\sqrt{2}}V_{cs}^*V_{ud}\left\{C_+\left[(\overline{u}d)_L(\overline{s}c)_L+(\overline{s}d)_L(\overline{u}c)_L\right]_{\mathbf{15}}\right.\nonumber\\
&&\qquad\qquad\qquad\qquad\left.+C_-\left[(\overline{u}d)_L(\overline{s}c)_L-(\overline{s}d)_L(\overline{u}c)_L\right]_{\mathbf{\overline{6}}}\right\},\nonumber\\
&&\mathcal{L}_{\mathrm{eff}}^{\mathrm{DCS}}=-\frac{G_F}{\sqrt{2}}V_{cd}^*V_{us}\left\{C_+\left[(\overline{u}s)_L(\overline{d}c)_L+(\overline{d}s)_L(\overline{u}c)_L\right]_{\mathbf{15}}\right.\nonumber\\
&&\qquad\qquad\qquad\qquad\left.+C_-\left[(\overline{u}s)_L(\overline{d}c)_L-(\overline{d}s)_L(\overline{u}c)_L\right]_{\mathbf{\overline{6}}}\right\}.\nonumber\\
&& \mathcal{L}_{\mathrm{eff}}^{\mathrm{SCS}}= -\frac{G_F}{\sqrt{2}}\frac{\lambda_s-\lambda_d}{2}\Big\{C_+\Big[(\overline{u}s)_L(\overline{s}c)_L+(\overline{s}s)_L(\overline{u}c)_L\nonumber\\
&&\qquad\qquad\qquad\qquad\qquad\quad -(\overline{d}d)_L(\overline{u}c)_L-(\overline{u}d)_L(\overline{d}c)_L\Big]_{\mathbf{15}}\nonumber\\
&&\qquad\qquad\qquad\qquad +C_-\Big[(\overline{u}s)_L(\overline{s}c)_L-(\overline{s}s)_L(\overline{u}c)_L\nonumber\\
&&\qquad\qquad\qquad\qquad\quad +(\overline{d}d)_L(\overline{u}c)_L-(\overline{u}d)_L(\overline{d}c)_L\Big]_{\mathbf{\overline{6}}}\Big\},\nonumber\\
&&\quad+\frac{G_F}{\sqrt{2}}\frac{\lambda_b}{4}\Big\{C_+\Big[(\overline{u}d)_L(\overline{d}c)_L+(\overline{d}d)_L(\overline{u}c)_L+(\overline{s}s)_L(\overline{u}c)_L\nonumber\\
&&\qquad\qquad\qquad\qquad+(\overline{u}s)_L(\overline{s}c)_L-2(\overline{u}u)_L(\overline{u}c)_L\Big]_{15}\nonumber\\
&&\qquad\qquad+C_+\sum_{q=u,d,s}\left[(\overline{u}q)_L(\overline{q}c)_L+(\overline{q}q)_L(\overline{u}c)_L\right]_{\mathbf{3}_+}\nonumber\\
&&\qquad\qquad+2C_-\sum_{q=d,s}\left[(\overline{u}q)_L(\overline{q}c)_L-(\overline{q}q)_L(\overline{u}c)_L\right]_{\mathbf{3}_-}\Big\}\;.
\end{eqnarray}
Here, 
we define $(\overline q_1q_2)_L\equiv\overline q_1\gamma_\mu(1-\gamma_5)q_2$. 
The subscripts on the square bracket  label the $SU(3)$ representations, with  
$3 \otimes 3\otimes \bar 3=\mathbf{15}\oplus\overline{\mathbf 6}\oplus \mathbf 3_+\oplus \mathbf 3_-$.
The  Wilson coefficients are written as $C_\pm=C_1\pm C_2$ since 
the  current--current operators $Q_1$ and $Q_2$ are more conveniently combined as $Q_\pm=(Q_1\pm Q_2)/2$,
where $Q_1=(\bar q_{1\alpha}c_{\beta})_L (\bar q_{3\beta}q_{2\alpha})_L$ and $Q_1=(\bar q_{1\alpha}c_{\alpha})_L (\bar q_{3\beta}q_{2\beta})_L$.
For SCS processes, the effective Lagrangian is   decomposed into two distinct parts: 
one proportional to $\lambda_s-\lambda_d$ and another proportional to $\lambda_b$.
For the latter, the SU(3) triplet representation originates from penguin operators, which have not been  shown  explicitly above and are conventionally expressed as $\lambda_b \sum\limits_{i=3}^{6} C_i O_i$~\cite{Buchalla:1995vs,Li:2012cfa}.
The $\overline{\mathbf{15}}$ representation arises from the interference of tree operators, using the identity $\lambda_d = -\lambda_s -\lambda_b$.
Therefore, the decay amplitudes can be categorized into a dominant  tree amplitude proportional to $\lambda_s$ and a smaller component proportional to $\lambda_b$, $M=\lambda_s A_s + \lambda_b A_b$.

Considering only the flavor structure,  five $SU(3)$  contraction can be constructed for  $H(15)$, 
but the K\"orner--Pati--Woo (KPW) theorem reduces them to a single independent term after imposing the quark color-symmetry constraints~\cite{Korner:1970xq,Pati:1970fg,Groote:2021pxt}.
By applying the KPW theorem and using the explicit forms of the initial and final states along with the effective Lagrangian, 
 the parity-violating amplitudes  $F$,
including penguin contributions, can be expressed in terms of invariant decay amplitudes as~\cite{Geng:2023pkr,He:2024unv}
% \begin{eqnarray}
% \mathcal{M}^{IRA}
% &=&a_{15} \times(T_{c\bar{3}})_i(H_{15})^{\{ik\}}_j(\overline{T_8})^j_kP^l_l\notag\\
% &+&b_{15} \times(T_{c\bar{3}})_i(H_{15})^{\{ik\}}_j(\overline{T_8})^l_kP^j_l\notag\\
% &+&c_{15} \times(T_{c\bar{3}})_i(H_{15})^{\{ik\}}_j(\overline{T_8})^j_lP^l_k\notag\\
% &+&d_{15} \times(T_{c\bar{3}})_i(H_{15})^{\{jk\}}_l(\overline{T_8})^l_jP^i_k\notag\\
% &+&e_{15} \times(T_{c\bar{3}})_i(H_{15})^{\{jk\}}_l(\overline{T_8})^i_j P^l_k\notag\\
% &+&a_{6} \times(T_{c\bar{3}})^{[ik]}(H_{\overline{6}})_{\{ij\}}
% (\overline{T_8})^j_kP^l_l\notag\\
% &+&b_{6} \times(T_{c\bar{3}})^{[ik]}(H_{\overline{6}})_{\{ij\}}(\overline{T_8})^l_kP^j_l\notag\\
% &+&c_{6} \times(T_{c\bar{3}})^{[ik]}(H_{\overline{6}})_{\{ij\}}(\overline{T_8})^j_lP^l_k\notag\\&+&d_{6} \times(T_{c\bar{3}})^{[lk]}(H_{\overline{6}})_{\{ij\}}(\overline{T_8})^i_kP^j_l.\notag\\
% &+&a_{3} \times(T_{c\bar{3}})_{i}(H_{3})^{k}(\overline{T_8})^i_l P^l_k\notag\\
% &+&b_{3} \times(T_{c\bar{3}})_{i}(H_{3})^{k}(\overline{T_8})^i_k P^l_l\notag\\
% &+&c_{3} \times(T_{c\bar{3}})_{i}(H_{3})^{i}(\overline{T_8})^k_l P^l_k\notag\\
% &+&d_{3} \times(T_{c\bar{3}})_{i}(H_{3})^{l}(\overline{T_8})^k_l P^i_k.
% \end{eqnarray}
\begin{eqnarray}\label{su3}
 \mathcal{F}^{IRA}&=&\tilde{f}^a(\eta_1)(\overline{\mathbf{B}})_k^j\mathcal{H}(\overline{\mathbf{6}})_{ij}(\mathbf{B}_c)^{ik}\notag\\ &+&\tilde{f}^b(P_8^\dagger)_k^l(\overline{\mathbf{B}})_j^k\mathcal{H}(\overline{\mathbf{6}})_{il}(\mathbf{B}_c)^{ij}\nonumber\\
 &+&\tilde{f}^c(P_8^\dagger)_j^l(\overline{\mathbf{B}})_l^k\mathcal{H}(\overline{\mathbf{6}})_{ik}(\mathbf{B}_c)^{ij}\nonumber\\
 &+&\tilde{f}^d(P_8^\dagger)_j^l(\overline{\mathbf{B}})_i^k\mathcal{H}(\overline{\mathbf{6}})_{kl}(\mathbf{B}_c)^{ij}\nonumber\\
 &+&\tilde{f}^e(P_8^\dagger)_k^l(\overline{\mathbf{B}})_j^i\mathcal{H}(\mathbf{15})_l^{jk}(\mathbf{B}_c)_i\nonumber\\
 %&+&\tilde{f}^f(\eta_1)(\overline{\mathbf{B}})_k^j\mathcal{H}(\mathbf{15})_j^{ik}(\mathbf{B}_c)_i\nonumber\\
 %&+&\tilde{f}^g(P_8^\dagger)_l^j(\overline{\mathbf{B}})_k^l\mathcal{H}(\mathbf{15})_j^{ik}(\mathbf{B}_c)_i\nonumber\\
 %&+&\tilde{f}^h(P_8^\dagger)_k^l(\overline{\mathbf{B}})_l^j\mathcal{H}(\mathbf{15})_j^{ik}(\mathbf{B}_c)_i\nonumber\\
 %&+&\tilde{f}^i(P_8^\dagger)_k^i(\overline{\mathbf{B}})_j^l\mathcal{H}(\mathbf{1}\mathbf{5})_l^{jk}(\mathbf{B}_c)_i\nonumber\\
 &+&\tilde{f}_3^a(\eta_1)(\overline{\mathbf{B}})_i^j\mathcal{H}(\mathbf{3})^i(\mathbf{B}_c)_j\nonumber\\
 &+&\tilde{f}_3^b(P_8^\dagger)_k^i(\overline{\mathbf{B}})_m^k\mathcal{H}(\mathbf{3})^m(\mathbf{B}_c)_i\nonumber\\
 &+&\tilde{f}_3^c(P_8^\dagger)_k^m(\overline{\mathbf{B}})_m^k\mathcal{H}(\mathbf{3})^i(\mathbf{B}_c)_i\nonumber\\
&+& \tilde{f}_3^d(P_8^\dagger)_n^m(\overline{\mathbf{B}})_m^k\mathcal{H}(\mathbf{3})^n(\mathbf{B}_c)_k\nonumber\\
&+&\tilde{f}_1^v(\mathbf{B}_c)^{il}\mathcal{H}(\overline{\mathbf{6}})_{ij}{\left(\overline{\mathbf{B}}_l^jP_k^m\hat{M}_m^k-\overline{\mathbf{B}}_l^kP_k^m\hat{M}_m^j\right)}\nonumber\\
&+&\tilde{f}_2^v(\mathbf{B}_c)^{il}\mathcal{H}(\overline{\mathbf{6}})_{ij}{\overline{\mathbf{B}}_m^jP_l^k\hat{M}_k^m}.\label{am}
\end{eqnarray}
The relevant symbols are summarized as follows: 
charmed baryon $\mathbf B_c$, final state baryon $\overline{\mathbf B}$, final state meson $P_8^\dagger/\eta_1$,
the effective weak Hamiltonian $\mathcal H(a)$ with representation $a=3,\bar 6, 15$,  where the repeated flavor indices $i,j,k,l,m,n=1,2,3$   is summed.
Each SU(3)-invariant amplitude $\tilde f$ is associated with an independent, a priori undetermined parameter.
The coefficients $\tilde{f}_3^a,\tilde{f}_3^b,\tilde{f}_3^c,\tilde{f}_3^d$ 
encode not only penguin contributions but also specific tree level terms.
The last two lines $\tilde f^v_{1,2}$ represent SU(3) breaking effects, dominated by the strange-quark condensate $\langle\bar s s\rangle$ with mass spurion $\hat M\propto\mathrm{diag}(0,0,1)$, as discussed in Ref.~\cite{Yang:2025orn}.  Later, we will find that these SU(3) breaking terms may play an important role in explaining existing data.
The parity-conserving  $G$
 amplitudes admit an analogous decomposition. 
\\

\noindent{\bf \textit{Decoding hierarchy and unexpected enhancement}}
From the  expansion in Eq.~(\ref{su3}), one obtains the decay amplitudes for individual channels  as summarized in Table~\ref{tab:SU(3)}.
Note that $A_s$ and $A_b$ denote the pure hadronic amplitudes without CKM factor, each given by a specific linear combination of the parameters $\tilde f$.
Multiplying $A_s $ by 
 $V_{ud}V_{cs}^*$, $\lambda_s$, and $V_{us}V_{cd}^*$ yields the CF, SCS, and DCS amplitudes, respectively, while  the $\lambda_bA_b$ term occurs only in the SCS sector, as shown in Table~\ref{tab:SU(3)}.
To isolate $A_b$, we establish relations among CF, DCS, and SCS  amplitudes.
%The tree-level $A_s$ contributes to all three types,  while $A_b$  acts only in SCS processes.
Once  $A_s$ is fixed from CF modes, $A_b$  can therefore be extracted from the SCS amplitudes.
%Using the  decay amplitudes in Eq.~\eqref{am}, we identify several representative decay processes, as summarized in Table~\ref{tab:SU(3)} according to their common $A_s$  structure.

\begin{table*}[htbp!]
\caption{ Representative decay channel with $SU(3)_F$  amplitudes $F$ and  experimental data. %Here $s_\theta=\sin \theta=0.22$ with $\theta$ Cabibbo angle. 
 The $G$ amplitudes can be written in a similar way  by replacing $f_i$ with $g_i$.}
\label{tab:SU(3)}\begin{tabular}{|c|c|c|c|c|c|c|c}\hline\hline
& channel & amplitude $F$ & SU(3) breaking $T_v$&  $Br_{exp}$ & $\alpha_{exp}$ \\\hline
\multirow{2}{*}{CF}&  $\Xi_c^0\to \Sigma^+K^-$ & $-\tilde f^c V_{ud}V_{cs}^*$ &   
& $0.18(4)\%$ & \cr\cline{2-6}
&$\Lambda_c^+\to \Xi^0 K^+$ & $\tilde f^c V_{ud}V_{cs}^*$ & $\tilde f_2^v V_{ud}V_{cs}^*$
&0.55(7)\%& 0.01(16)\\\hline
\multirow{2}{*}{ SCS}& $\Xi_c^0\to \Sigma^+\pi^-$ & $\tilde f^c \lambda_s + \lambda_b (\tilde f_3^b+\tilde f_3^c+\tilde f^c/2)$ & &&\cr\cline{2-6}
& $ \Xi_c^0\to pK^-$ & $-\tilde f^c \lambda_s + \lambda_b (\tilde f_3^b+\tilde f_3^c-\tilde f^c/2)$ & & & \\\hline
\multirow{2}{*}{DCS}& $\Xi_c^0\to p\pi^-$ & $V_{us}V_{cd}^*\tilde f^c$ & & & \cr\cline{2-6}
& $\Xi_c^+ \to n\pi^+$ &$-V_{us}V_{cd}^*\tilde f^c$ &&&
\\\hline\hline
CF & $\Xi_c^0\to \Xi^-\pi^+$ & $(\tilde f^e-\tilde f^b)V_{ud}V_{cs}^*$ & &  $1.43(27) \%$& $-0.640(51)$  \\\hline
\multirow{2}{*}{ SCS}& $\Xi_c^0\to \Sigma^-\pi^+$&$-(\tilde f^e-\tilde f^b)\lambda_s+\lambda_b(\tilde f^c_3+\tilde f_3^d-\frac{\tilde f^e}{4}+\frac{\tilde f^b-\tilde f^e}{2})$& &  & \cr\cline{2-6}
& $ \Xi_c^0\to \Xi^-K^+$& $(\tilde f^e-\tilde f^b)\lambda_s+\lambda_b(\tilde f^c_3+\tilde f_3^d-\frac{\tilde f^e}{4}-\frac{\tilde f^b-\tilde f^e}{2})$ & $\tilde f_1^v (\lambda_s+\lambda_b/2)$ &$0.039 (11)\%$&\\\hline
DCS& $\Xi_c^0\to \Sigma^-K^+$& $-V_{us}V_{cd}^*(\tilde f^e-\tilde f^b)$ & $-\tilde f_1^v V_{us}V_{cd}^*$ & & \\\hline\hline
CF& $\Xi_c^0\to \Xi^0 \pi^0$ & $\frac{(\tilde f^b-\tilde f^d)}{\sqrt{2}}V_{ud}V^*_{cs}$ & &  0.69(14)\% & $-0.90(28)$ \\\hline
\multirow{2}{*}{ SCS}& $\Lambda_c^+\to \Sigma^+ K_S$ &  $\frac{(\tilde f^b-\tilde f^d)}{\sqrt{2}}\lambda_s+ \lambda_b\frac{\tilde f_3^b+(\tilde f^b-\tilde f^d)/2}{\sqrt{2}}$ & $-\frac{\tilde f_1^v}{\sqrt{2}}(\lambda_s+\lambda_b/2)$  & 0.048(14)\% &\cr\cline{2-6}
& $\Lambda_c^+\to \Sigma^0K^+$ &$\frac{(\tilde f^b-\tilde f^d)}{\sqrt{2}}\lambda_s+ \lambda_b\frac{\tilde f_3^b+(\tilde f^b-\tilde f^d)/2} {\sqrt{2}}$& $-\frac{\tilde f_1^v}{\sqrt{2}}(\lambda_s+\lambda_b/2)$ & 0.0373(31)\% & $-0.54(20)$\\\hline
\hline
\end{tabular}
\end{table*}

To elucidate how our  methods decodes $A_b$ amplitudes using Table~\ref{tab:SU(3)},
we first express the decay amplitudes $F$ in terms of the CF, SCS and DCS ones, given by 
\begin{eqnarray}
 && F(CF)=V_{ud}V_{cs}^*\; ( T^f_s+T_{v,C}^f)=V_{ud}V_{cs}^*\;  A^f_{s,C} ,\; \notag\\
 && F(SCS)=\sigma \lambda_s  \; T^f_s + \lambda_b \; T^f_b + T_{v,S}^f(\lambda_s+\lambda_b/2) \notag\\
 &&\qquad\qquad=\sigma \lambda_s  A^f_{s,S} + \lambda_b A^f_{b,S} \;,\quad \sigma=\pm 1\;, \notag\\
&&  F(DCS)=V_{us}V_{cd}^*\; (T^f_s+T^f_{v,D})=V_{us}V_{cd}^*\; A^f_{s,D},\label{amc}
\end{eqnarray}
where we define $A^f_{s,S}=T^f_s+ \sigma T^f_{v,S}$ and $A^f_{b,S}= T^f_b+ T_{v,S}^f/2$ with the subscript $C,S,D$ serving as abbreviations  of CF, SCS and DCS,  respectively.
Here $T_s (T_b)$ is the pure hadronic amplitude without CKM factor $\lambda_s (\lambda_b)$ as listed in the third column of Table~\ref{tab:SU(3)}, while $T_v$ denotes the dominant $SU(3)$ breaking correction in the fourth column.
In the absence of such specification, the general notations $A_s^f$ and $A_b^f$ are employed.
And the sign of $\sigma=\pm 1$ depends on the specific CF and SCS decay chains.
Each amplitude  $A_{s,b}$  represents specific linear combinations of the reduced amplitudes $\tilde f$  listed in Table~\ref{tab:SU(3)}. 
The amplitude $A^f_{s}$ contains only tree-level contributions, whereas  $A^f_{b}$ receives both tree-level and penguin contributions. 
The $G$ amplitudes can be written similarly, with  $A^f$ replaced by $A^g$.

The ratios of the squared amplitudes between SCS and  CF (or DCS) channels can then be expressed as
\begin{eqnarray}
  &&  \frac{|M(SCS)|^2}{|M(CF)|^2}=\frac{|V_{us}|^2}{|V_{ud}|^2}\left(1+2\sigma\mathrm{Re}(R^P_T)+O(\lambda_b^2)\right)\;,\nonumber\\
  && \frac{|M(SCS)|^2}{|M(DCS)|^2}=\frac{|V_{cs}|^2}{|V_{cd}|^2}\left(1+2\sigma\mathrm{Re}(R^P_T)+O(\lambda_b^2)\right)\;,
\end{eqnarray}
where% $|M|^2$ and $R^P_T$ are
\begin{eqnarray}\label{RPT}
&&\!\! \!\!  \!\!  \!\! \! 
R^P_T=\frac{ \left(A^f_{s,C/D}+\frac{\sigma}{2} A_{v}\right)A_{v}^*
+\frac{ \lambda^*_b}{ \lambda_s^*} A^f_{s,S} A^{f*}_{b,S} + \kappa^2 (f\to g)}{|A^f_{s,C/D}|^2 + \kappa^2 |A^g_{s,C/D}|^2},\notag
\end{eqnarray}
we define $A_{v}=A_{s,S}-\sigma A_{s,C/D}$.
Here the dimensionless  quantity $R_T^P$ quantities the deviation of  SCS amplitude from its SU(3)-related CF or DCS reference.
Note that $R^P_T$ receives two distinct contributions,
 SU(3) breaking term 
$A_v$ and $\lambda_b/\lambda_s$ suppressed term.
Since amplitude ratios are directly related to branching fractions and other decay observables, $R^P_T$ can be expressed in terms of the corresponding branching-fraction ratio as
\begin{eqnarray}
 && \mathrm{Re}(R^P_T)=\frac{1}{2\sigma}\left(a\; r_a-1\right)
 =\frac{1}{2\sigma}\left(c \;r_c-1\right),
 \\
 &&\text{with}\;  a=\frac{K(CF)|V_{ud}|^2}{K(SCS)|V_{us}|^2},\; 
  c=\frac{K(DCS)|V_{cd}|^2}{K(SCS)|V_{cs}|^2}
 \;,\notag\\
  && K(X)=\tau_{B_c} %\frac{\sqrt{\lambda(m_{B_c}^2,m_B^2,m_P^2)}}{2m_{B_c}} 
  p_B((1+m_B/m_{B_c})^2-m_P^2/m_{B_c}^2)
  \;.\notag\label{BRR}
\end{eqnarray}
where we define $r_a=Br(SCS)/Br(CF)$ and $r_c=Br(SCS)/Br(DCS)$.
Here  $\tau_{B_c}$ and $K(X)$ denote the lifetime of  charmed baryon, and  phase-space factor for the decay channel $X$, respectively. We also define  the factors $a,c$ to explain the differences in  phase-space and  CKM factors  between the paired modes.
Since $A_b$ also contains  tree amplitudes, it is natural to expect $A_s$ and $A_b$ to be of the same order, $A_b/A_s=\mathcal O(1)$. This implies
  $R^P_T\sim T_v/A_{s,C}+\lambda_b/\lambda_s$, so that 
the value $R^P_T$  is controlled by the interplay between SU(3) breaking effects and the CKM-suppressed ratio $\lambda_b/\lambda_s$.
In the exact symmetry limit $T_v\to 0$, it reduces to  $R^P_T\sim 10^{-3}$.

From Table~\ref{tab:SU(3)}, 
 three paired CF–SCS branching fractions have been measured experimentally in Ref.~\cite{Belle:2024ikp,PDG}, yielding the following ratios as
\begin{eqnarray}\label{anomaly}
  &&\mathrm{Re}(R^P_T)_{exp}\Big|^{\Xi_c^0\to \Xi^-K^+}_{\Xi_c^0\to \Xi^-\pi^+}%=   \frac{1}{2}\left(a\frac{Br(SCS)}{Br(CF)}-1\right)
  =-0.21(10),\notag\\
 &&\mathrm{Re}(R^P_T)_{exp}\Big|^{\Lambda_c^+\to \Sigma^+ K_S}_{\Xi_c^0\to \Xi^0\pi^0}%=   \frac{1}{2}\left(a\frac{Br(SCS)}{Br(CF)}-1\right)
   =0.10(21)\;, \nonumber\\
   &&\mathrm{Re} (R^P_T)_{exp}\Big|^{\Lambda_c^+\to \Sigma^0 K^+}_{\Xi_c^0\to \Xi^0\pi^0} % =\frac{1}{2}\left(a\frac{Br(SCS)}{Br(CF)}-1\right)
   =-0.03(10)\;, 
\end{eqnarray}
where the superscript and subscript  on $Re(R^P_T)$  denote SCS and CF, respectively.   
The extracted central values are of $\mathcal{O}(0.1)$, about two orders of magnitude above the naive expectation $\sim 10^{-3}$ in the exact SU(3) symmetry limit discussed above.

In the limit of exact SU(3) symmetry,  the absence of $A_b$ causes the ratio  vanish.
However, 
for $\Xi_c^0\to \Xi^-\pi^+$ and $\Xi_c^0\to \Xi^-K^+$, 
the extracted ratio departs from zero at the $2.1\sigma$ level, indicating a non-negligible contribution from
 $\lambda_b A_b$.  
A naive data analysis indicates  $Re(R^P_T)$ can reach $\mathcal{O}(10\%)$, implying a comparable CPV effect. 
Such a value requires the underlying hadronic ratio,
$Re(e^{i\delta_{CP}}(A^{f*}_b/A^{f*}_s +\kappa^2 A^g_s A^{g*}_b/ |A^f_s|^2)( 1 + \kappa^2 |A^g_s/A^f_s|^2))$ 
to be of order $\mathcal{O}(100)$,   roughly two orders of magnitude above  the naive expectation of $\mathcal{O}(1)$, after factoring out the CKM factor  $\lambda_s/\lambda_b$. Here    $\delta_{CP}$ is the relative weak phase  between $\lambda_s$ and $\lambda_b$. 
This points to an unexpected enhancement in these decays.

Additionally, the Lee-Yang parameters provide an independent handle on the extraction of the $A_b$ amplitude~\cite{Lee:1957qs}. For CF and DCS processes, we have
\begin{eqnarray}\label{br}
  &&
   \alpha_{CF,DCS}=\frac{2\kappa\mathrm{Re}(F^*G)}{|F|^2+\kappa^2|G|^2}
   = \frac{2\kappa\mathrm{Re}(A^{f*}_{s,C/D} A^g_{s,C/D})}{|A^f_{s,C/D}|^2+\kappa^2|A^g_{s,C/D}|^2},
   \nonumber\\
  % &&\beta_{CF, DCS}=\frac{2\kappa\mathrm{Im}(F^*G)}{|F|^2+\kappa^2|G|^2}
  % = \frac{2\kappa\mathrm{Im}(A^{f*}_{s,C/D} A^g_{s,C/D})}{|A^f_{s,C/D}|^2+\kappa^2|A^g_{s,C/D}|^2}\notag\\
  % &&\gamma_{CF, DCS}=\frac{|F|^2-\kappa^2|G|^2}{|F|^2+\kappa^2|G|^2}
  % = \frac{|A^f_{s,C/D}|^2-\kappa^2|A^g_{s,C/D}|^2}{|A^f_{s,C/D}|^2+\kappa^2|A^g_{s,C/D}|^2}.\notag\\
\end{eqnarray}
Analogous expressions for $\beta_{CF, DCS}$ and $\gamma_{CF, DCS}$
are obtained by replacing the numerator $\mathrm{Re}(F^*G)$ in $\alpha$ with 
 $\mathrm{Im}(F^*G)$ and $|F|^2-\kappa^2|G|^2$, respectively.
And the CKM factors contained in $F,G$ cancel  in the numerator and denominator.

For  SCS corresponding processes, we would have
\begin{eqnarray}
  &&\alpha_{SCS}= \frac{2\kappa\mathrm{Re}((A^{f*}_{s,S}+(z A^f_{b,S})^*)(A^g_{s,S} + z A^g_{b,S}))}{|A^f_{s,S}+z A^f_{b,S}|^2+\kappa^2|A^g_{s,S} + z A^g_{b,S}|^2},\nonumber\\
  % &&\beta_{SCS}=\frac{2\kappa\mathrm{Im}((A^{f*}_{s,S}+(z A^f_{b,S})^*)(A^g_{s,S} + z A^g_{b,S}))}{|A^f_{s,S}+z A^f_{b,S}|^2+\kappa^2|A^g_{s,S} + z A^g_{b,S}|^2},\nonumber\\
  % &&\gamma_{SCS}=\frac{|A^f_{s,S}+z A^f_{b,S}|^2-\kappa^2|A^g_{s,S} +z A^g_{b,S}|^2}{|A^f_{s,S}+z A^f_{b,S}|^2+\kappa^2|A^g_{s,S} + z A^g_{b,S}|^2}.
\end{eqnarray}
where we define $z=\sigma\lambda_b/\lambda_s$.
Analogous expressions hold for $\beta_{SCS}$ and $\gamma_{SCS}$.
Note that these observables provide independent sensitivity to the $A_b$ contribution.
They are parametrized by the relative deviations 
$X_{SCS}-X_{CF/DCS}$, where  $X = \alpha, \beta, \gamma$. These ratios  can be expressed by
\begin{eqnarray}
&&\alpha_{SCS}-\alpha_{CF/DCS}=2{\rm Re} (R^P_T(\alpha))+O(\lambda_b^2),\\
&&\beta_{SCS}-\beta_{CF/DCS}= 2{\rm Im}(R^P_T(\beta))+O(\lambda_b^2),\notag\\
&&\gamma_{SCS}-\gamma_{CF/DCS}=-4 { \rm  Re}(R^P_T(\gamma))+O(\lambda_b^2),\notag
\end{eqnarray}
with the components as
\begin{eqnarray}
&& R^P_T(\alpha)=\frac{1}{2}\left(\alpha_{SCS}|_{z\to 0}-\alpha_{CF/DCS}\right)\notag\\
&&\qquad + \kappa z\frac{(A^f_{s,S} A^g_{b,S}- A^f_{b,S} A^g_{s,S})((A^{f*}_{s,S})^2-\kappa^2(A^{g*}_{s,S})^2)}{ (|A^f_{s,S}|^2 + \kappa^2 |A^g_{s,S}|^2)^2},\notag\\
&& R^P_T(\beta)=\frac{1}{2}\left(\beta_{SCS}|_{z\to 0}-\beta_{CF/DCS}\right)\notag\\
&&\qquad+\kappa z \frac{(A^{f*}_{s,S})^2+\kappa^2(A^{g^*}_{s,S})^2}{(|A^f_{s,S}|^2+\kappa^2|A^g_{s,S}|^2)^2} (A^f_{s,S}A^g_{b,S}-A^g_{s,S}A^f_{b,S}),\notag\\
&& R^P_T(\gamma) = -\frac{1}{4}\left(\gamma_{SCS}|_{z\to 0}-\gamma_{CF/DCS}\right)\notag\\
&&\qquad+\kappa^2 z \frac{A^{f*}_{s,S} A^{g*}_{s,S}(A^f_{s,S} A^g_{b,S}-A^g_{s,S} A^f_{b,S})}{ (|A^f_{s,S}|^2 + \kappa^2 |A^g_{s,S}|^2)^2}.
\end{eqnarray}
Note that the first terms in $R^P_T(\alpha,\beta,\gamma)$ arise from the breaking effects, and therefore vanish in the exact SU(3) symmetry limit.
As an illustration, 
consider the paired CF–SCS system, $\Xi_c^0 \to \Xi^0 \pi^0$ and $\Lambda_c^+ \to \Sigma^0 K^+$, where the measured $\alpha$ values are $-0.90(28)$ and $-0.54(20)$, respectively.
This suggests 
\begin{eqnarray}
    \alpha(\Lambda_c^+ \to \Sigma^0 K^+)- \alpha(\Xi^0_c \to \Xi^0 \pi^0) =0.36 \pm 0.34\;.
\end{eqnarray}
In the exact SU(3) symmetry, 
it  exhibits a similarly large enhancement of the hadronic amplitude albeit within 1$\sigma$ level, consistent with  the pattern identified in  $R^P_T$. A systematic investigation of these observables therefore provides an  independent and complementary probe.
\\

\noindent{\bf \textit{Possible explanation}}
Such an enhancement is difficult to attribute to ordinary QCD dynamics alone. It may still reflect the limited precision of current data, but it could also point to additional underlying dynamics. If confirmed with higher accuracy, it would call for a nontrivial explanation, most notably enhanced SU(3) flavor breaking or new physics with additional CPV phases.

We first examine whether the observed enhancement can be accommodated within SU(3) breaking. As shown in Table~\ref{tab:SU(3)}, all three channel pairs in Eq.~\ref{anomaly} receive sizable SU(3) breaking corrections. Using the central values of the global fit parameters from Ref.~\cite{Yang:2025orn} together with Eq.~\ref{RPT},  we predict $Re(R^P_T)=(-0.13,-0.14,-0.14)$ for the three pairs,  respectively. 
Similarly, we predict $Re(R^P_T(\alpha))=-0.08$ between $\Lambda_c^+ \to \Sigma^0 K^+$ and $\Xi^0_c \to \Xi^0 \pi^0$.
 These values remain consistent with current experimental results within uncertainties,  suggesting that
 SU(3) breaking provides a viable explanation of  the observed enhancement.  
  In particular, a typical breaking size of $T_v/A_s \sim 10\%$ is already sufficient to reproduce the data, without requiring a large intrinsic enhancement of $A_b/A_s$.

To determine whether the enhancement truly originates from dominant SU(3) breaking effects in Ref.~\cite{Yang:2025orn}, it is essential to study decay modes that are relatively insensitive to such corrections.
From Table~\ref{tab:SU(3)},  several CF-SCS paired channels 
satisfy this criterion,
 including $\Xi_c^0\to \Sigma^+K^--\Xi_c^0\to \Sigma^+\pi^-$, $\Xi_c^0\to \Sigma^+K^--\Xi_c^0\to pK^-$, $\Xi_c^0\to \Xi^-\pi^+-\Xi_c^0\to\Sigma^-\pi^+$.
We focus on CF-SCS pairs because DCS modes are experimentally more challenging. Currently, no data are available for these paired channels, highlighting the need for dedicated measurements.
Among them,  $\Xi_c^0\to \Sigma^+\pi^-$ and $\Xi_c^0\to pK^-$ are particularly promising and may serve as   golden channels for  CPV searches.
Their $W$-exchange topology suppresses the tree amplitude $\lambda_s A_s$~\cite{Cheng:1993gf}, enhancing the relative impact of $\lambda_b A_b$ terms to make sizable CPV feasible.
Observation of a similar enhancement  in 
modes insensitive to these dominant SU(3) breaking effects would provide a clean test of its dynamical origin. 
Its presence would strongly favor new physics,  whereas its absence would  point to SU(3) breaking effects as the dominant explanation.
\\

%If future data confirm the enhancement while disfavoring an SU(3) breaking origin, the result would provide a clear indication of new physics. In that case, the decay amplitude may be modified as  $M\to M+e^{i\delta_{NP}}A_{NP}$, where the new contribution is not CKM suppressed by $\lambda_b$ and can naturally account for the observed enhancement.

\noindent{\bf \textit{Conclusion} }

 We have developed an SU(3)-based, data-driven strategy to disentangle the $\lambda_s A_s$ and $\lambda_b A_b$ amplitudes in SCS charmed-baryon decays. Current data indicate an enhancement of $A_b/A_s$ over the naive O(1) expectation at the $2.1\sigma$ level, which can be accommodated by $\mathcal{O}(10\%)$ SU(3) breaking but may also signal new physics. Lee-Yang parameters provide an independent test. In particular, the SU(3)-insensitive modes $\Xi_c^0\to \Sigma^+\pi^-$ and $\Xi_c^0\to pK^-$, where the tree contribution is suppressed, offer clean discriminants between the two possibilities. Precise measurements of their branching fractions and Lee-Yang parameters can therefore directly probe CP violation in the charm-baryon sector.
\\

\noindent{\bf \textit{Acknowledgments} }
The work of Zhi-Peng Xing is supported by NSFC under grant   No. 12405113. 
The work of Jin Sun is supported by IBS under the project code, IBS-R018-D1. 
 Xiao-Gang He was supported by the Fundamental Research
Funds for the Central Universities, by the National Natural Science Foundation of the People’s
Republic of China (No.12090064, 12205063, 12375088 and W2441004).

\bibliographystyle{JHEP}
\bibliography{main}

\end{document}